\documentstyle[11pt,gh2001-asp,twoside,epsf]{article}
\begin{document}
\title{Constraints on nested bars -- implications for gas inflow}
\author{Witold Maciejewski}
\affil{Osservatorio Astrofisico di Arcetri, Largo Fermi 5, 50125 Firenze,
Italy, and Jagiellonian University Observatory, Cracow, Poland} 

\begin{abstract}
A wide-spread belief that nested bars enhance gas inflow to the galactic 
centre has recently been contradicted by dynamical models in which inner 
bars seem to prohibit such inflow. Can the existing models of dynamically 
possible double bars be modified to enable strong inflow in the secondary 
bar? I present here simple dynamical arguments which imply that in general, 
double bars in resonant coupling do not enhance gas inflow. However, stronger
inflow with straight shocks in the inner bar can occur if there is no
resonant coupling of the commonly assumed form between the bars.
\end{abstract}

\section{Introduction}
Any non-axisymmetric perturbation in mass distribution in the galactic disk 
exerts torques, which pull the disk matter out of circular orbits. The 
resulting trajectories,
which often intersect, can be populated by stars, but gas clouds will collide
and drop down the potential well. Thus in general, an asymmetric gravitational
potential enhances gas inflow towards its central parts.

The most common asymmetry in the disk plane developed in unstable disks
is an $m=2$ bar-like mode. Various studies of gas flow in
barred galaxies (see Athanassoula 2000 for a recent review) indicate
that the gas transport {\it into} the central kpc
differs from that {\it within} the central kpc. On large scale,
two straight shocks develop on the leading edges of the bar, and gas falls
towards the center along these shocks. Nevertheless, if the galaxy has an
inner Lindblad resonance (ILR), the inflow may stagnate inside of it, with
gas settling on the nuclear ring there. This mode of gas flow has importance 
for star-forming nuclear rings observed in barred galaxies. 
Recently, it has been found that when the cloud velocity 
dispersion is high enough, gas may avoid stagnating on the nuclear ring,
and can flow inwards along a nuclear spiral (Englmaier \& Shlosman 2000).

If another, smaller bar exists inside the ILR of the large-scale bar, and if
gas flow in this secondary bar is analogous to the one in the main bar, then
this secondary bar may force gas deeper into the potential well. This scenario
is a modification of the fueling mechanism proposed by Shlosman et al. 
(1989), which originally involved gaseous bars. Rapidly increasing number of 
galaxies in which nested stellar bars have been seen (Laine et al. 2002)
brought attention to this scenario. However, Maciejewski \& Sparke (2000)
showed that such double bars are unlikely to increase the inflow because of 
their orbital structure. Here I examine assumptions behind this last claim, 
and explore how to waive them.
%whether, and how, it is possible to waive them.

\section{The standard model of Maciejewski \& Sparke (2000)}
Since the two bars in doubly barred galaxies can be seen virtually at any 
relative orientation, they most likely rotate independently, each with 
its own pattern speed. Such double bars will not admit any closed periodic 
orbits in general, which served as a backbone of the steady potential of a 
single bar. In order to find regular stellar orbits supporting double bars, 
Maciejewski \& Sparke (2000) assigned sets of test particles to imaginary 
closed strings that change shape as the bars rotate through each other, but 
return to their original positions after the two bars realign. They call these
sets of particles {\it loops}, and if the loop remains aligned with one bar or
the other, then it contributes to the backbone of this bar. Maciejewski \& 
Sparke constructed a potential which admits simultaneously loops following 
the inner and the outer bar. Regular (though not closed) orbits of particles
populating these loops support the potential of their model, making it 
dynamically possible. This is the standard model of Maciejewski \& 
Sparke.

In order to minimize the number of chaotic zones around resonances,
the standard model assumes resonant coupling between the bars: 
corotation (CR) of the inner bar coincides with the ILR of the outer
bar. In this potential, loops originating from the $x_1$ orbits in the 
large-scale bar (the $x_1$ loops) continue to support that bar, while loops 
that come from the $x_2$ orbits (orthogonal to the large-scale bar, and
extending inwards from its ILR), now 
diversify. The outer $x_2$ loops remain perpendicular to the outer bar, 
but the inner ones start to follow the secondary bar in its motion,
forming its backbone. Since the dynamics of the outer $x_2$ loops is 
still dominated by the outer bar, the dynamically possible inner bar 
cannot extend that far in radius, and should end well inside the ILR 
of the outer bar. In resonant coupling, 
this means that the secondary bar should end well inside its own CR.

Only a bar extending almost to its CR develops straight shocks 
(Athanassoula 1992) . If the bar is shorter, shocks
start to curve and weaken, and eventually disappear into a ring around the 
bar. Because the inner bar in the standard model extends only to about half 
of its CR radius, it may lack principal shocks or dust lanes.
In addition, loops that support it originate from the 
$x_2$ orbits in the outer bar: they are rather round, with no cusps, so 
there is no reason for shocks in the gas flow to develop. Recently,
Maciejewski et al. (2002) modeled hydrodynamically gas inflow in the 
standard model, and confirmed predictions from the orbital analysis: in 
their models no stationary straight shocks
develop in the secondary bar, but the flow organizes into various elliptical 
and circular rings. 

\begin{figure}
\plotfiddle{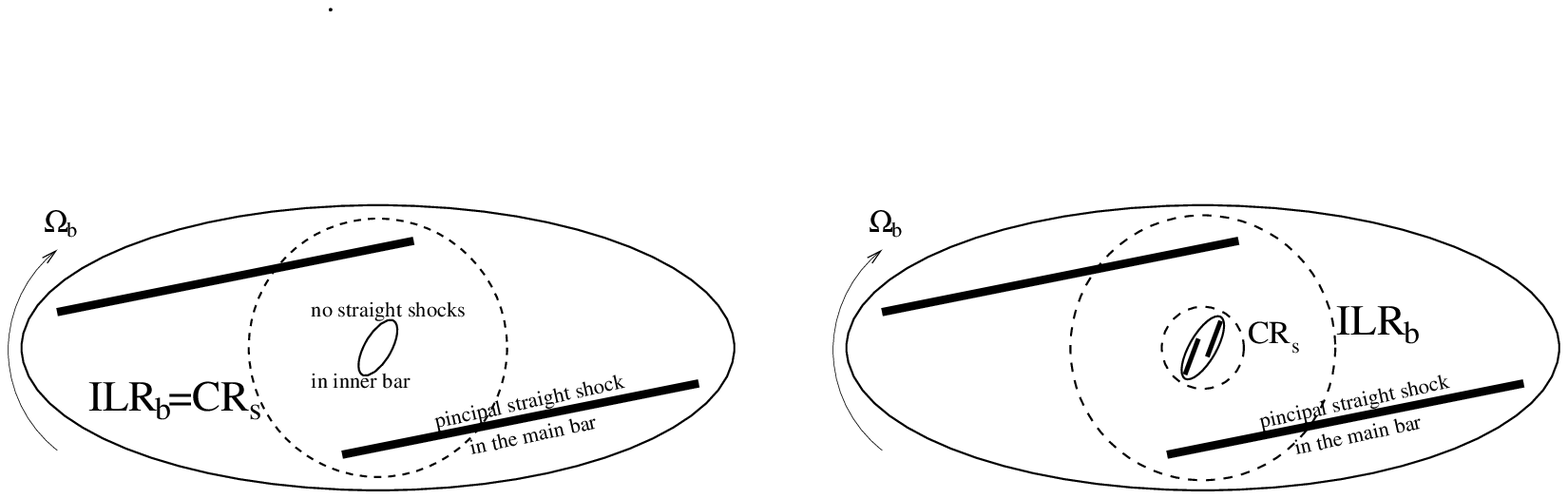}{4cm}{0}{80}{80}{-230}{-390}
\vspace{-1cm}
\caption{Two alternative setups of nested bars outlined by solid ellipses.
CR of the inner bar (CR$_S$), and ILR of the outer bar (ILR$_B$) are marked 
with dashed circles. {\it Left:} bars in resonant coupling with the inner bar 
ending well inside its own CR. {\it Right:} no resonant coupling between the 
bars --- each bar can extend to its own CR, generating inflow in its own pair 
of shocks.}
\end{figure}

\section{Proposed alterations to the standard model}
In the standard model, the inner bar does not extend to its CR 
radius, and therefore the gas flow that it induces is completely
different from that in the outer bar. Two assumptions of the standard model 
may influence this result. First, this model was constructed to explore
possibly biggest secondary bars, and has a large radius of outer bar's ILR
(38\% of its CR radius). Secondly, the resonant coupling was assumed. 
Below, I examine what happens when these two assumptions are lifted.

\underline{\bf Resonant coupling with smaller ILR radius:} 
The ILR can be moved to arbitrarily small radii by
decreasing the extent of the central mass concentration. For example,
a power-law rotation curve $\Omega = a r^{-b}$ implies the ratio of the ILR and
CR radii to be $(1- \sqrt{1-(b/2)})^{1/b}$, which can be 
arbitrarily small for $b \rightarrow 0$ (i.e. when approaching solid body 
rotation). Now, consider a large-scale bar extending almost to its CR, 
and the inner bar ending at about the ILR of that large-scale bar. If the
ILR/CR ratio is small, potential of the inner bar should dominate
the region inside the ILR, and the inner bar should be able to drag all the
$x_2$ loops. In this case, a doubly barred system in resonant coupling
with both bars extending to their respective CRs would be possible.

Although only loop calculations can verify plausibility of this model,
its setup is unrealistic if the secondary bar forms 
from the instability in the nuclear disk in the outer bar. Simulations of
gas flow in a single bar show that the nuclear ring or disk forms when the
ILR is present, with a family of $x_2$ orbits inside it. Gas moving into
these $x_2$ orbits is responsible for the offset of the straight shocks
from the bar's major axis {\it already inside} the ILR, and outside of
the nuclear ring (disk, spiral), on which gas eventually settles. 
Thus the outer radius of the nuclear disk is always considerably smaller
than the ILR radius. If the inner bar forms from this nuclear disk, 
there is no disk material to extend the inner bar to its CR, 
which for resonant coupling is at the outer bar's ILR. This reaffirms 
findings of the standard model: the inner bar cannot extend to its own 
CR in doubly barred galaxies with resonant coupling. Accordingly,
resonant coupling excludes straight shocks in the inner bar, implying 
that such double bars {\it do not} increase gas inflow like the large-scale 
bars do. The left panel of Figure 1 illustrates this situation.

\underline{\bf No standard resonant coupling:}
An alternative to the standard model is shown in the right panel of
Figure 1, where the standard resonant coupling between the bars has been 
lifted. In principle, the secondary bar in such systems can extend to its 
CR radius, and it can create shocks in gas flow, which in turn 
enhance the inflow. However, systems without any resonant coupling are
likely to be unstable, because the number of chaotic zones doubles there.

One can consider a different resonant coupling when
the $\Omega-\kappa/2$ curve has a local maximum, outside of which it 
drops to zero at very small and large radii. In this case, the line of 
constant bar pattern speed intersects this curve twice, and {\it two} 
ILRs form. The standard case was assuming
resonant coupling between the outer ILR of the large bar and the 
CR of the small bar. Now, consider coupling between
the {\it inner} ILR of the large bar and the small bar's CR.
The dynamics of such a system 
drastically differs from the standard model, because inside the inner ILR 
there are no $x_2$ orbits, from which the loops supporting the inner
bar used to originate. Detailed loop calculations may find how 
such systems support themselves, but there are no clear candidates
that can naturally follow the secondary bar. In addition, some inner ILRs 
observed at the small bar's end can be artificial effects of data resolution:
seeing or beam smearing makes inner velocity growth look linear,
which falsely implies an inner ILR.
% (see the discussion for NGC 4303 in Schinnerer et al. 2002). 
The lack of an inner ILR can in fact be beneficial for the inner bar,
because in this case the $x_2$ loops that support it may extend all the
way to the galactic centre.

\section{Conclusions}
By constructing a dynamically possible doubly barred galaxy, Maciejewski
\& Sparke (2000) showed that double bars, just like single bars, can be 
supported by regular orbits. However, unlike single bars, their inner bar
does not extend to its CR radius, and therefore it neither creates
shocks in gas flow, nor increases the inflow. The loop approach was
necessary in reaching this conclusion, because it allows to determine the 
dynamics of the interface between bars. 

Contrary to the standard model of Maciejewski \& Sparke, a wide-spread
conviction persists that nested bars should increase gas inflow. Here, 
I attempted to waive two 
most vulnerable assumptions of the standard model in order to find 
candidates of dynamical systems, in which the secondary bar could extend
to its CR, therefore being able to enhance the inflow. 
I found that making the ILR radius of the large-scale bar smaller can
lead to a self-consistent model of a doubly barred galaxy in resonant
coupling with both bars extending to their CRs, but such a model 
is unrealistic if the secondary bar forms from 
the instability in the nuclear disk. Both bars extending to their
CRs may occur in systems without any resonant coupling, but
such systems may prove chaotic. One should expect completely different
dynamics when the CR of the small bar couples with the inner ILR 
of the large bar.

\end{document}